\documentstyle[aps,multicol,epsfig]{revtex}

\def\be{\begin{equation}}          \def\ee{\end{equation}}
\def\bea{\begin{eqnarray}}         \def\eea{\end{eqnarray}}
\def\q{\quad}       
\def\I{\hat{\vec I}}
\def\hI{{\hat I}}   \def\hT{{\hat T}}  \def\hB{{\hat B}}
\def\ms{m^{\prime}}       \def\on{\omega_0}
\def\re#1{Eq.~(\ref{#1})}
\def\av#1{\langle{#1}\rangle}
\def\tp{\tau_+}       \def\tm{\tau_-}
\begin{document}

\title{Nuclear Spin Relaxation for Higher Spin}

\author{W. Apel$^{1}$ and Yu.~A. Bychkov$^{2,1}$ }

\address{$^{1}$ Physikalisch-Technische Bundesanstalt,
Bundesallee 100, 38116 Braunschweig, Germany.}

\address{$^{2}$ L.D. Landau Institute for Theoretical Physics,
ul.Kosigina, 2, Moscow, Russia.}

\date{\today}

\maketitle

\begin{abstract}
We study the relaxation of a spin $I$ that is weakly coupled to a quantum
mechanical environment.
Starting from the microscopic description, we derive a system of coupled
relaxation equations within the adiabatic approximation.
These are valid for arbitrary $I$ and also for a general stationary
non--equilibrium state of the environment.
In the case of equilibrium, the stationary solution of the equations becomes
the correct Boltzmannian equilibrium distribution for given spin $I$.
The relaxation towards the stationary solution is characterized by a set of
relaxation times, the longest of which can be shorter, by a factor of up to
$2I$, than the relaxation time in the corresponding Bloch equations
calculated in the standard perturbative way.
\end{abstract}

\begin{multicols}{2}

Nuclear magnetic resonance is a well--established method for testing
electronic properties in solids \cite{Slichter}.
In recent years, it became possible to apply this technique not only in
three dimensions, but also to a two--dimensional electron system,
the quantum Hall ferromagnet that is realized in semiconductor
heterostructures in a strong magnetic field.
The experimental work lead to the unexpected conclusion that a new kind of
low--energy states, Skyrmions, can be formed and can determine the nuclear
relaxation processes in these systems when one Landau sub--level of one spin
direction is filled
\cite{DKSWP88,BDGK90,BDPWT95,TBDPW95,KKBPW98,KKBPW98b,MFHBLBS00,%
LesH:Girvin,CMBFGS97,SGJM00}.

As a theoretical description of spin relaxation, Bloch's equations have been
successfully used for about fifty years now.
While these phenomenological equations are applicable in a wide range of cases,
their microscopic derivation reveals two main restrictions.
First, as was already discussed in the original work \cite{WB53},
the derivation becomes strictly valid, if either the spin is $I=1/2$
or the temperature of the bath is large compared to the resonance frequency.
But the spin in the system under study can be $I=3/2$
(for ${}^{69}$Ga, ${}^{71}$Ga, and ${}^{75}$As; cf.~Ref.~\cite{DKSWP88}),
or higher in the case of magnetic impurities.
Further, the progress in the experimental techniques now lets a regime of
temperatures and magnetic fields come into reach, in which the temperature
of the bath is of the same order as the nuclear resonance energy
(nuclear Zeeman energy).
The second restriction in the derivation of the phenomenological equations
demands that the environment (bath), the quantum mechanical degrees of freedom
causing the spin relaxation, be in thermodynamic equilibrium.
But in the case of the quantum Hall ferromagnet, the nuclear spins are coupled
to a two--dimensional electron gas in which the electron--electron interaction
plays the key role, since all single particle states are degenerate into a
single Landau level due to the strong magnetic field; such a system,
dominated by Skyrmion--states, is not necessarily in equilibrium.
Thus, it appears worthwhile to reconsider the microscopic derivation of
phenomenological equations for the spin relaxation in order to investigate
whether there is a significant difference between the general case and a
case in which the Bloch equations are valid.

It is the purpose of this work to relax the two above restrictions
by investigating the general case of an arbitrary spin $I$ and also an
arbitrary stationary state of the environment responsible for the relaxation.
We refrain from studying a specific mechanism and consider instead the general
case of a magnetic moment coupled to a bath of other quantum degrees of freedom.
This magnetic moment can be a nuclear spin, or also a magnetic impurity.
In the following, we use the terms ``nuclear spin'' for the magnetic moment
and ``electrons'' for the bath - usually the latter is called ``lattice''.
Then, the contribution of the nuclear spin to the Hamiltonian is

\be
H = - \gamma \I  \cdot ( {\vec B}_0 + {\hat {\vec B}} )  \;. \label{ha}
\ee

Here, the magnetic moment, $\gamma \I$ (where $\I$ is the spin),
couples to an effective magnetic field.
It is well known that for $I > 1/2$, there is an additional term causing
relaxation, the electric quadrupole moment of the nucleus coupled to an
inhomogeneous external electric field.
Here, we have omitted this term in $H$, since the model \re{ha} already
suffices for the demonstration of our method; an inclusion of a quadrupolar
coupling is straightforward.
The effective magnetic field acting on the nuclear spin $\I$ in \re{ha}
contains an operator $\hat {\vec B}$ of the electronic system.
One can picture $\hat {\vec B}$ as being proportional to the electrons'
spin.
Its longitudinal component, $\hat B_z$, modifies the eigenvalues of the
nuclear spin system, while the transverse components cause transitions between
eigenstates.
There is also a fixed part of the magnetic field in $z$--direction,
${\vec B}_0 = B_0 \hat e_z$, which acts as an external field.
The coupling beween nuclear spin and electrons is supposed to be weak in the
sense that we can use the adiabatic approximation as discussed below.
We do not need to make any assumptions about the electronic subsystem's
Hamiltonian or the electronic subsystem's state.
This Hamiltonian may contain electron--electron interactions and the subsystem
may be in an arbitrary stationary state, equilibrium or non--equilibrium.

We want to derive kinetic equations for the expectation value of the spin
vector $\I$ ($\I{}^{2} = I(I+1)$).
To this end, we need to take all components of the density matrix into account,
not just only the spin vector.
Then, as will be seen below, it is advantageous to use the spherical
tensor operators $\hT_{LM}$ as a complete basis in the space of operators
acting on the state of a spin $I$.
The $\hT_{LM}$ are the irreducible tensor operators in the spherical
coordinate representation \cite{Rose}.
For actual calculations, the following definition using the spherical harmonics
$Y_{LM}$ proves very helpful:

\be
\hT_{LM} \;:=\; {\cal N}_{IL} \;\; (\I  \cdot {\vec \nabla})^L \;\;\;
          r^{L} \; Y_{LM}(\hat r)   \;\;.       \label{telm}
\ee

$\hT_{LM}$ is a polynomial in the components of the spin operator.
It is independent of the auxiliary variable $\vec r$,
since on the r.h.s., there are $L$ derivatives acting on a polynomial of order
$L$ ($\hat r = \vec r /r$ denotes the unit vector; we use the conventions of
Ref.~\cite{Rose} for the spherical harmonics $Y_{LM}$).
For a spin $I$, the $(2I+1)^2$ operators $\hT_{LM}$ with
$L=0 \cdots 2I$ and $M=-L \cdots L$ form a complete system of operators acting
in the spinspace.
The normalization, ${\cal N}_{IL} = 2^L \sqrt{4\pi(2I-L)!/(2I+L+1)!} / L!$,
is choosen such that
$
 Sp \left\{ \; \hT^{\dagger}_{LM} \;\; \hT_{L^{\prime}\,M^{\prime}} \right\}
    \;=\;  \delta_{L\;L^{\prime}} \;\; \delta_{M\;M^{\prime}} \; ,
$
and we have $\hT^{\dagger}_{LM} = (-1)^M \hT_{L,-M}$.
Specific expressions for the operators $\hT_{L,M}$ (for $L=0\cdots 4$)
can be found in Table IV of Ref.~\cite{AES62}.

After having established the basic notation, we proceed now to describe the
derivation of the kinetic equation for the average $T_{LM}(t) = \av{\hT_{LM}}$,
to find the stationary solution and to study finally the relaxation
towards the stationary solution.
Here and below, the brackets $\av{\cdots}$ stand for the state of the combined
system of spin and electrons.
We use the framework of the Keldysh method \cite{K65} in order to derive
the kinetic equation.
As in our earlier work on the electron spin relaxation \cite{AB99},
we shall employ the adiabatic approximation; that means that in the equation
of motion, the effect of the coupling between spin and electrons is neglected
beyond the first order in the spin's eigenenergies and also neglected beyond
the second order in the relaxation times,
cf.~Ref.~\cite{Landau-Lifshitz-V10P91}.
The unperturbed motion of the spin is a precession with the frequency
$\on = \gamma \, B_0$.
Then, we have up to the second order of the perturbation theory in the coupling
$\hat V = - \gamma \I \cdot {\hat {\vec B}}$:

\bea
&& \left( i\;  \partial_t \;-\; M\; \on \right) \; T_{LM}(t) =  \nonumber \\
&& \;\; \av{\left[ \hT_{LM}(t) , \hat V(t) \right] }            \nonumber \\
&& -i \; \int^t_{-\infty} dt_1 \; \av{\left[ \left[ \hT_{LM}(t) , \hat V(t)
  \right], \hat V(t_1) \right] }  \;+\; {\cal O}(\hat V^3)   \;.   \label{kg}
\eea

In the spirit of the adiabatic approximation, we now consider the terms of the
perturbation series in higher than second order as giving rise to an
additional (weak) time dependence of the spin operators due to the
relaxation process, and we decouple the expectation values of spin and electron
operators.
The first order term,
$\av{[\hT_{LM},\I](t)}\cdot \av{{\hat {\vec B}}(t)}$,
describes the Knight shift, the shift of the nuclear resonance frequency
due to the coupling to the electrons.
Since we want to focus on the relaxation, we disregard this corrections of
the spin's eigenfrequency in the following.
Then, we get

\bea
&& \left( i\;  \partial_t \;-\; M\; \omega_0 \right) \;
   T_{LM}(t) =  \nonumber \\
&& -i \sum_{m;\ms=0,\pm 1}
   \left( \av{\left[ \left[\hT_{LM}, \hI_{m}\right], \hI_{\ms} \right](t) }
    \;\;C_{-m;-\ms} \right. \nonumber \\
&& \mbox{\rule{1cm}{0mm}}
    + \left. \av{\left\{ \left[\hT_{LM}, \hI_{m}\right], \hI_{\ms} \right\}(t) }
    \;\;R_{-m;-\ms}  \right)                       \label{kg1}
\eea

In deriving \re{kg1}, we took the unperturbed time dependence of the operator
$\I(t_1)$ into account, but neglected the difference $t-t_1$ in its time
dependence due to the relaxation.
In \re{kg1}, equal--time commutators ($[,]$) and anticommutators ($\{,\}$) can
be evaluated, and the expectation values can again be expressed by $T_{LM}(t)$
as we shall see below.
Thus, it is the use of the spherical tensor operators that makes it possible to
derive closed coupled equations for the relaxation of the spin.
Most crucial are the averages of the electronic subsystem that enter \re{kg1}.
Since we do not assume thermodynamic equilibrium for the electronic subsystem,
we get both, a correlation function and also a response function which are
independent and which we denote by $C_{m;\ms}$ and $R_{m;\ms}$, respectively.
They are given by

\bea
C_{m;\ms} &=& \frac{\gamma^2}{2} \int\limits_{-\infty}^t \!\! dt_1
    e^{-i\ms\on(t-t_1)}
    \av{\left\{ \hB_m(t), \hB_{\ms}(t_1) \right\} }       \\
R_{m;\ms} &=& \frac{\gamma^2}{2} \int\limits_{-\infty}^t \!\! dt_1
    e^{-i\ms\on(t-t_1)}
    \av{\left[ \hB_m(t), \hB_{\ms}(t_1) \right] }
\eea

Our convention for the vector components of $\I$ and ${\hat {\vec B}}$ is:
$\hI_{\pm1}=\hI_x \pm i \hI_y$, $\hI_0=\hI_z$ and
$\hB_{\pm1}=(\hB_x \pm i \hB_y)/2$, $\hB_0=\hB_z$.

Commutators and anticommutators in \re{kg1} can be calculated either
directly from the definition \re{telm}, or with the aid of the theory
of spherical tensors (cf.~Ref.~\cite{R62}).
The commutators express the behavior of the $\hT_{LM}$ under rotations:

\begin{mathletters}   \label{kt}
\bea
\left[ \hT_{LM}, \hI_z \right] &=& -M\;\hT_{LM} \\
\left[ \hT_{LM}, \hI_{\pm} \right] &=&
 -\sqrt{L(L+1)-M(M\pm1)} \;\;\hT_{L,M\pm 1}  \;.
\eea
\end{mathletters}

The anticommutators can also be expanded in $\hT_{LM}$;
specializing the general result containing Racah coefficients
(see Eq.~(17) of ref.~\cite{R62}) to our case, we get for $m=0,\pm1$:

\be
  \left\{ \hT_{LM}, \hI_m \right\} \,=\,
   a^{(m)}_{L+1;M} \; \hT_{L+1,M+m}
   \,+\, b^{(m)}_{L;M} \; \hT_{L-1,M+m}         \label{akt}
\ee

with the coefficients
\bea
&& a^{(0)}_{L;M} \;=\; b^{(0)}_{L;M} \;=\; \sqrt{ L^2-M^2 } \; c_L \;, \\
&& a^{(\pm 1)}_{L;M} = b^{(\mp 1)}_{L;M\pm 1} =
  \mp \sqrt{ (L+1\pm M)(L\pm M) } \; c_L \;,
\eea
where
\be
  c_{L} \;=\; \sqrt{ \frac{(2I+1)^2 - L^2}{(2L)^2-1} }  \;.
\ee

Inserting relations (\ref{kt}) and (\ref{akt}) into \re{kg1} solves our task of
deriving a closed set of relaxation equations for arbitrary nuclear spin $I$.
The equations are linear in the expectation values of the spherical tensor
operators $\hT_{LM}$.
The properties of the electronic system enter the equations parametrically
in the form of correlation functions and response functions.
Since the spin vector is given by the $\hT_{1M}$, a relaxation equation
for $\av{\I}$ can be extracted from the system of equations (\ref{kg1}).
In the case of $I=1/2$, this is particularly simple, since $\hT_{LM}=0$ for
$L>1$ (and $\hT_{00}=1/\sqrt{2}$ is constant) and we recover the Bloch
equations, see below.

In this general form, the equations (\ref{kg1}) are still not very transparent.
Therefore, we now make additional assumptions regarding the correlation and
response functions entering \re{kg1}.
The term $\hI_z \, \hB_z$ in the perturbation $\hat V$ (corresponding to
$m, \ms =0$ in \re{kg1}) just changes the spin's resonance frequency $\on$.
Since we already omitted the Knight shift, the first order correction term
to this frequency, we now also omit consistently the second order
contributions resulting from $m=0$ or $\ms =0$ in \re{kg1}.
Next, the terms with $m=\ms$ in \re{kg1} are neglected too.
In the electron system, they would correspond to a total change of the
$z$--component of the spin by $2$; if the electrons' state is a strict
eigenstate to $z$--component of the spin, such expectation values vanish.
Under these assumptions, both correlation and response functions,
$C_{m;\ms}$ and $R_{m;\ms}$, now become ``diagonal'' in the index $m$,
$C_{m;\ms}=\delta_{m;-\ms} C_m$ and $R_{m;\ms}=\delta_{m;-\ms} R_m$.
Now, from the general structure of \re{kg1}, it is obvious that the
equations do not couple $T_{LM}(t)$ for different $M$.
So finally, we arrive at our general result, valid for arbitrary spin $I$ and
for a non--equilibrium state of the bath

\bea
 &&\left(  \partial_t \;+\;i\,M\, \on \right) \; T_{LM}(t) \;=\; \nonumber \\
 && - \frac{1}{2 \tp}  \; \left[L(L+1)-M^2\right] \;\; T_{LM}(t) \nonumber \\
 && - \frac{1}{2 \tm} \;
   \left\{ L\sqrt{(L+1)^2-M^2} \; c_{L+1} \;\; T_{L+1,M}(t) \right. \nonumber\\
 && \mbox{\hspace{5ex}} \left.
   \;-\; (L+1)\sqrt{L^2-M^2} \; c_{L} \;\; T_{L-1,M}(t) \right\}  \; .
 \label{res}
\eea

We have defined two different times by
$1/\tp := 2(C_{1} + C_{-1})$ and $1/\tm := 2(R_{1} - R_{-1})$.
These times are independent, as long as we assume a non--equilibrium state
of the electronic system.
We have $\tp \ge 0$ and the ratio $\tp /\tm$ can be expressed as
$\tp /\tm = (1-q)/(1+q)$ with positive $q$.
Thus, $| \tp/\tm | \le 1$.
If the electrons are in equilibrium at temperature $T$,
the fluctuation--dissipation theorem results in $q=\exp(- \on /T)$.

We want to stress that these relaxation equations are valid (under the
assumptions stated above), no matter which specific relaxation mechanism one
wants to consider.
The special mechanism enters the equations in the form of two time scales,
$\tp$ and $\tm$.
In the case of thermodynamic equilibrium in the bath, their ratio, $\tp / \tm$,
is fixed by the temperature, and only the time $\tp$
-- which also depends on temperature --
is specific for the relaxation mechanism.

Relaxation equations generally serve two purposes.
First, they determine a stationary state.
Second, they describe the relaxation towards this state.
We now want to discuss both of these points.

{\it Stationary solution}\\
The study of the stationary solution of \re{res} serves as an important test of
our procedure, since, in the case of equilibrium in the bath, the result is
obvious.
With the ansatz $T_{LM}(t) = \delta_{M;0}\;T_L$, one derives a recursion
relation

\be
    c_{L+1} \;\; T_{L+1} \;=\; - \frac{\tm}{\tp}  \;\; T_{L}
  + c_{L} \;\; T_{L-1}   \;\;, \;\; L = 1 \cdots 2I
\ee

where $T_0 = (2I+1)^{-1/2}$, $T_{2I+1}=0$.
The solution is
\be
 I_z \;=\; \frac{2}{3}\,I(I+1) \; (1-q) \; {\cal M}_{21}/{\cal M}_{11}
\ee
where ${\cal M}_{ij}$ are elements of the matrix
${\cal M}=M_1 \cdot M_2 \cdots M_{2I}$ with
\be
 M_L \;=\;  {1+q \q (1-q)^2 \; c_{L+1}^2 \choose 1 \q\q\q 0 \q\q }  \; .
\ee
Evaluation of this matrix product yields -- for a general non--equilibrium
state of the bath -- the following stationary distribution for the
$z$--component of the spin vector

\be
  I_z \;=\;  \sum_{m=-I}^I m \; q^{-m}  \;/\; \sum_{m=-I}^I q^{-m} \;\;.
\ee

If we assume now the electronic system is in equilibrium at temperature $T$,
we find that $I_z$ is given by the Brillouin function, the well--known correct
equilibrium distribution for a spin $I$.

{\it Relaxation}\\
The general solution of the coupled relaxation equations \re{res} is
a superposition of exponentially decaying terms.
We are mostly interested in the relaxation of the spin components
$\hI_z \propto \hT_{10}$ and $(\hI_x + i \hI_y) \propto \hT_{11}$, which are
described by the part $M=0$ and $M=1$ of the full system of equations.
For $M=0$ and $M=1$, there are $2I$ different relaxation times,
for $M=2$ there are $2I-1$ times; in total we have $I(2I+3)$ times.
Here, each case of $I$ needs to be discussed separately.

We start with spin $1/2$.
In this case, the \re{res} gives immediately the standard Bloch equations
\cite{Slichter}:

\bea
 &&  \partial_t  \; I_z(t) \;=\; - \frac{1}{\tp} \;
            \left( I_z(t) \; - \frac{\tp}{2\tm} \right) \nonumber \\
 &&\left(  \partial_t \;+\;i \on \right) \; I_+(t) \;=\;
             - \frac{1}{2\tp} \; I_+(t)
\eea

\begin{figure}[thb]
\begin{center}
\epsfig{file=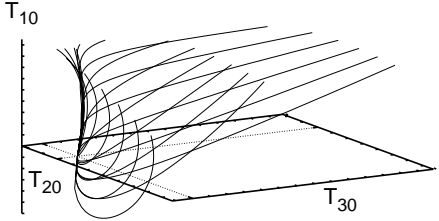,width=6cm}
\end{center}
\caption{Spin $I=3/2$: Solution of the relaxation equations for $M=0$ with
$\tp /\tm =0.8$; $I_z \propto T_{10}$ (arbitrary units used).}
\label{dreid}
\end{figure}
\begin{figure}[thb]
\begin{center}
\epsfig{file=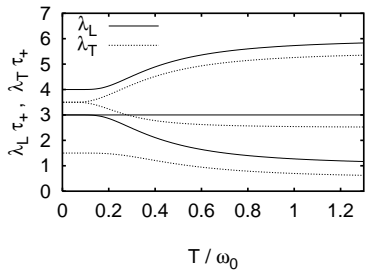,width=5cm}
\end{center}
\caption{Spin $I=3/2$: Inverse relaxation times in the longitudinal
(solid line) and transverse (dashed line) relaxation equations in units
of $1/\tp$.}
\label{lambda-long}
\end{figure}

More interesting, because of its experimental relevance \cite{DKSWP88}, is the
case of spin $3/2$.
The relaxation equations for $M=0$ couple the expectation values of three
operators: $\hT_{10}$, $\hT_{20}$, and $\hT_{30}$.
These equations are easily solved.
In Fig.~\ref{dreid}, we show the flow of the expectation values towards the
stationary state at the origin in the case of $\tp /\tm =0.8$.
The inverse time scales, $\lambda$, which determine the relaxation are shown
for both the longitudinal ($\lambda_L$, $M=0$) and the transverse
equations ($\lambda_T$, $M=1$) in units of $1/\tp$ as functions of the
temperature in the case of equilibrium in the bath in Fig.~\ref{lambda-long}.
Between the high--temperature limit $T>>\on$ and the low--temperature limit
$T<<\on$, both the largest longitudinal and the largest transverse relaxation
time decrease by a factor of three as compared to the relaxation time in
Bloch approximation calculated in the usual perturbative way.

For general $I$, we determine both the longitudinal and the transverse
relaxation times in the following limiting cases.
For $T>>\on$ ($\tp /\tm \sim 0$), the longitudinal relaxation time is
$\tp$ and the transverse $2\tp$ as in the Bloch equations.
In the opposite case, $T<<\on$ ($\tp /\tm \to 1$), all the operators
$\hT_{LM}$ become equally important in \re{res}.
Then, the $2I$ relaxation times are as follows ($n=0,1,\cdots 2I-1$):
In the longitudinal case, $M=0$, we get $\tp /[2(n+1)I-n(n+1)]$ and the longest
relaxation time ($n=0$) is always twofold degenerate;
in the transverse case, $M=1$, we get $\tp /[(2n+1)I-n^2]$ and here, the
longest relaxation time ($n=0$) is always non--degenerate.
These explicite expressions are a conjecture based on an explicite evaluation
of the relaxation equations for spin $I$ up to $7/2$.

Starting from the microscopic description, we have derived relaxation equations
valid for an arbitrary spin $I$ coupled to a quantum mechanical environment
in a non--equilibrium state.
The solution of these coupled equations shows that, compared to the Bloch
equations, there is an additional temperature dependence in the relaxation time
which can decrease the relaxation time by a factor of up to $2I$.

We would like to thank the Deutsche Forschungsgemeinschaft
for its support (Ap 47/3-1, 436 RUS 17/73/00).
Yu.~A.B.~wishes to acknowledge support from grants
RFFI-00-02-17292, IR-97-0076, and INTAS 99-01146,
and wishes to thank the PTB, where this work was performed, for its
hospitality.

\end{multicols}
\end{document}